\documentclass[aip,rsi,reprint,graphicx]{revtex4-2} 

\usepackage{graphicx}
\usepackage{subcaption}

\draft 

\begin{document}


\title{Using convolutional neural networks to detect edge localized modes in DIII-D from Doppler backscattering measurements} 



\author{N.Q.X. Teo}
\email[]{NATH0020@e.ntu.edu.sg}
\affiliation{Institute of High Performance Computing (IHPC), Agency for Science, Technology and Research (A*STAR), Singapore 138632, Singapore}
\affiliation{School of Physical and Mathematical Sciences, Nanyang Technological University, Singapore 637371, Singapore}

\author{V.H. Hall-Chen}
\affiliation{Institute of High Performance Computing (IHPC), Agency for Science, Technology and Research (A*STAR), Singapore 138632, Singapore}

\author{K. Barada}
\affiliation{Department of Physics and Astronomy, University of California, Los Angeles, California 90095, USA}

\author{R.J.H. Ng}
\affiliation{Institute of High Performance Computing (IHPC), Agency for Science, Technology and Research (A*STAR), Singapore 138632, Singapore}

\author{L. Gu}
\affiliation{AIP, RIKEN, Tokyo 103-0027, Japan}

\author{A.K. Yeoh}
\affiliation{Department of Physics, University of Oxford, Oxford
OX1 3PU, UK}
\affiliation{Institute of High Performance Computing (IHPC), Agency for Science, Technology and Research (A*STAR), Singapore 138632, Singapore}

\author{Q.T. Pratt}
\affiliation{Department of Physics and Astronomy, University of California, Los Angeles, California 90095, USA}

\author{X. Garbet}
\affiliation{School of Physical and Mathematical Sciences, Nanyang Technological University, Singapore 637371, Singapore}

\author{T.L. Rhodes}
\affiliation{Department of Physics and Astronomy, University of California, Los Angeles, California 90095, USA}


\date{\today}

\begin{abstract}
In H-mode tokamak plasmas, the plasma is sometimes ejected beyond the edge transport barrier. These events are known as edge localized modes (ELMs). ELMs cause a loss of energy and damage the vessel walls. Understanding the physics of ELMs and by extension, how to detect and mitigate them, is an important challenge. In this paper, we focus on two diagnostic methods --- Deuterium-alpha (D$_\alpha$) spectroscopy and Doppler backscattering (DBS). The former detects ELMs by measuring Balmer alpha emission while the latter uses microwave radiation to probe the plasma. DBS has the advantage of having higher temporal resolution and robustness to damage. These advantages of DBS diagnostics may be beneficial for future operational tokamaks and thus data processing techniques for DBS should be developed in preparation. In sight of this, we explore the training of neural networks to detect ELMs from DBS data, using D$_\alpha$ data as the ground truth. With shots found in the DIII-D database, the model is trained to classify each time step based on the occurrence of an ELM event. The results are promising. When tested on shots similar to those used for training, the model is capable of consistently achieving a high f1-score of 0.93. This score is a performance metric for imbalanced datasets that ranges between 0 and 1. We evaluate the performance of our neural network on a variety of ELMs in different high confinement regimes (grassy ELM, RMP mitigated, and wide-pedestal) finding broad applicability. Beyond ELMs, our work demonstrates the wider feasibility of applying neural networks to data from DBS diagnostics.
\end{abstract}


\maketitle 
\section{Introduction}
Edge localized modes (ELM) are events that occur in H-mode tokamak plasmas where plasma is ejected beyond the edge transport barrier \cite{zohm_edge_1996}. This results in a loss of energy and particles, and can cause damage to the tokamak walls. Diagnostics to detect these events are thus important for current and future tokamaks \cite{leonard1999impact}. Currently, ELMs can be detected using Deuterium-alpha (D$_\alpha$) spectroscopy \cite{heidbrink_hydrogenic_2004}, where the Balmer alpha emission from charge exchange reactions is measured. While not traditionally the primary diagnostic for ELM detection, ELM signatures have been observed by the Doppler backscattering (DBS) \cite{hirsch_doppler_2001, burrell2016discovery, ponomarenko2023investigation, yashin2023determination} diagnostic which uses microwave radiation to probe the plasma. DBS has the advantage of having higher temporal resolution \cite{chowdhury_novel_2023} and robustness to damage \cite{volpe_prospects_2017}, which may be beneficial for future burning plasmas. The quasi-optics and the longer wavelengths of microwave frequency allow for better resilience to deposition and radiation damage \cite{orsitto2016diagnostics}, and reduced sensitivity to mechanical vibrations \cite{varavin2019study}. However, detecting ELMs from DBS measurements is not straightforward and is labor intensive; as such, there is a need to develop automated tools. This paper explores the use of convolutional neural networks (CNN) to detect ELMs from DBS data. A model is trained on DBS data from multiple DIII-D discharges while using D$_\alpha$ data as the ground truth, tasked with classifying each time step based on the occurrence of an ELM. Similar work has been done using beam emission spectroscopy (BES) instead of DBS, where a deep neural network was trained to predict the likelihood of ELM crashes \cite{BES}. BES, however, functions in the optical frequencies, and thus may be prone to damage \cite{BES_disadvantages}. Moreover, it is unlikely that neutral beam injection will be in future tokamaks as the torque they inject creates instabilities in large machines \cite{creely2023sparc, chrystal2020predicting}. 

The aim of this research is to act as a proof-of-concept to demonstrate the ability of neural networks to detect ELMs from DBS data. Ultimately, the goal is to build tools for detection and prediction of ELMs in the burning plasmas of the next generation of tokamaks.

\section{Data} 
All data used in this research was obtained from the DIII-D \cite{Buttery:DIIID:2023} database.

\begin{figure*}
\includegraphics[scale=0.32]{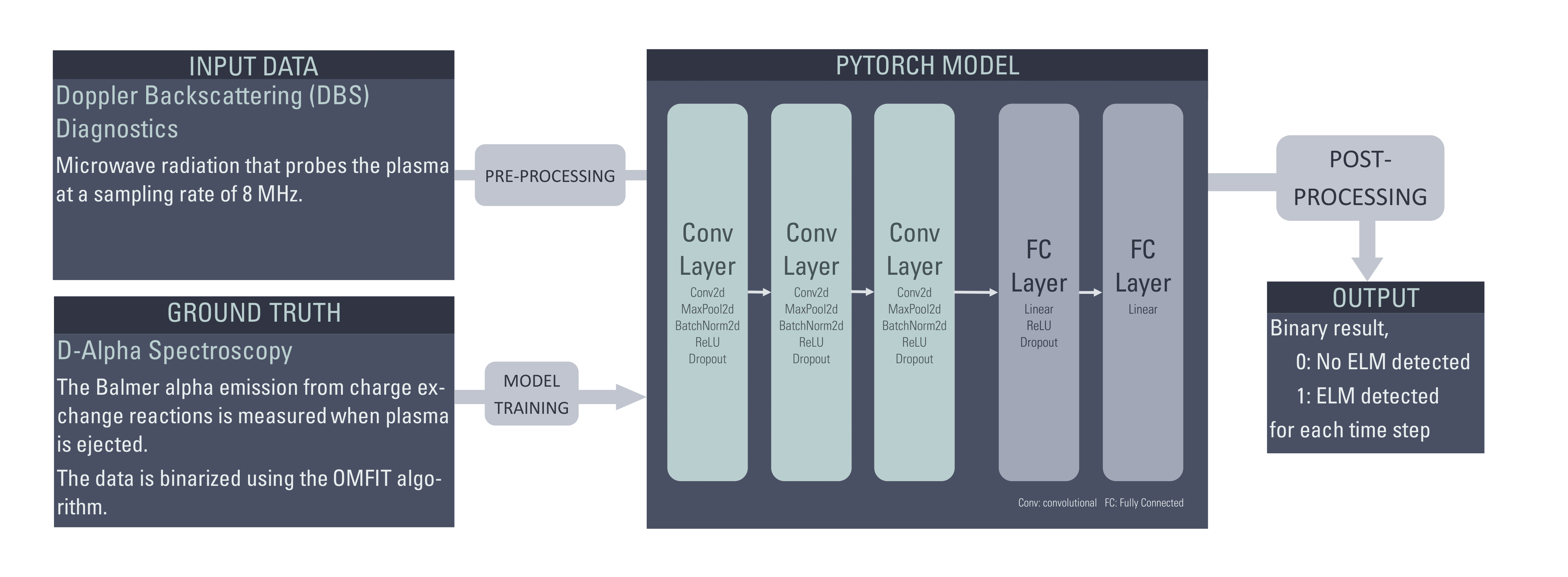}
\caption{Diagram of our neural network architecture. The model takes DBS data as input and outputs a binary ELM classification for each time step and is trained using data from D$_\alpha$ spectroscopy as the ground truth.}
\label{nn_diagram}
\end{figure*}
 
\subsection{Input Data}
Raw DBS data is a time-series waveform with real and complex components from eight channels, each operating at a different microwave frequency \cite{peebles2010novel}. Channels with higher probe frequencies penetrate further into the plasma, measuring perturbations at different depths within the pedestal. Only three channels were used for training and testing, channels 1 (55 GHz), 4 (62.5 GHz), and 7 (72.5 GHz), to minimize model size. This data was preprocessed before being fed to the model. The preprocessing involves performing a discrete fast Fourier transform to the waveform to obtain a power spectrum density (PSD), where each time step contains data of the signal power as a function of Doppler shifted frequency. We then smoothed the data by performing a running mean over the frequency domain of each time step as the model learned faster after smoothing. The preprocessed data is a 3D array with dimensions of time, channel (probe beam frequency), and Doppler shifted frequency.

\subsection{Ground Truth}
The ground truth is the target that the model trains to replicate. In this study, D$_\alpha$ measurements of the shots used were taken as the ground truth. The raw data was binarized using the ELM module in OMFIT \cite{meneghini2013integrated, meneghini2015integrated}, which calls a peak detection script. The output was then resampled to match the time steps of the input data. The final result is a 1D time-series array with 0 representing no ELM event and 1 representing an ongoing ELM event.

\section{Model}
\subsection{Model Architecture}
The model is a binary classifier built in the Python library, PyTorch \cite{NEURIPS2019_9015}, that classifies each time step of a given DBS PSD (Figure \ref{nn_diagram}). To classify a single time step, the model takes a window of DBS PSD data as input. The window contains DBS data of the current and previous time steps, spanning 128 time steps in total (2 ms), acting as a form of `memory'. The window size is chosen to be roughly the duration of the broadband spike of an ELM event.

The model contains two main sections --- convolutional blocks and a classifier as shown in Figure \ref{nn_diagram}. As the data is simple, the model was designed to be shallow. The data is first fed to three convolutional blocks of size 32, 64 and 128, containing the following PyTorch layers in order --- Conv2d, MaxPool2d, BatchNorm2d, ReLU, and Dropout (Figure \ref{nn_diagram}). The Conv2d layer performs a 2D convolution over the PSD window for each of the three channels. MaxPool2d downsamples the data and the ReLU (Rectified Linear Unit) activation function adds non-linearity to the model. BatchNorm2d and Dropout are functions to reduce overfitting, improving generalization of the model. The classifier comprises of the follwing PyTorch layers in order --- Linear, ReLU, Dropout, Linear. The Linear function is a layer where all input and output neurons are linearly connected. Both linear layers are of size 128. The logit output (inverse sigmoid output) passes through a sigmoid function and a rounding function to produce the final binary result. The general model design and structure for this convolutional neural network has been successfully implemented in other research \cite{nguyen2020imu} and is hence replicated for this project. 

\subsection{Training}
To train our model, we used DBS and D$_\alpha$ data from three shots --- 170869, 170870, and 170878 --- which were part of an experimental effort to understand the inter-ELM turbulence behavior \cite{barada2021new} and were recorded at a sampling rate of 8 MHz, as opposed to the routine 5 MHz. These discharges have input power very close to the L-H power threshold and hence have nearly evenly spaced large type-I ELMs.  

BCEWithLogitsLoss was used as the loss function; it uses binary cross entropy to calculate the loss between the model output and ground truth. Optimization was carried out using the Adam optimizer, incorporating weight decay for regularization. The Adam optimizer is an algorithm that allows for efficient model training. Additionally, the ReduceLROnPlateau scheduler was implemented to dynamically adjust the learning rate in response to stagnant loss values.

Given that the data is in a time-series format and thus inherently ordered, there was no implementation of data shuffling. All training shots were concatenated, and the resultant training dataset was split such that the final 20\% of the dataset served as the validation set. Without shuffling, the validation set is biased and hence further testing was done on independent shots to assess the model performance.

\subsection{Testing} \label{Testing}
Initial testing was performed on three shots. These shots contain type I ELMs and were recorded at a sampling rate of 8 MHz. Of these, two are notable and will be discussed, namely 170877 and 170880. The model was also tested on other shots to ascertain its generality. These were either recorded at different sampling rates or contain different ELM types. 

Further processing was applied to the model output to improve the model performance. This involved using the running mean of the sigmoid output (output from the sigmoid function applied to the logits) to mitigate small timescale anomalies that are too fast to be consistent with ELM patterns. All parameters used were kept constant for testing to avoid inflating the model performance.

\section{Results}
\begin{figure*}
\centering
\begin{subfigure}{0.9\columnwidth}
    \includegraphics[width=\textwidth]{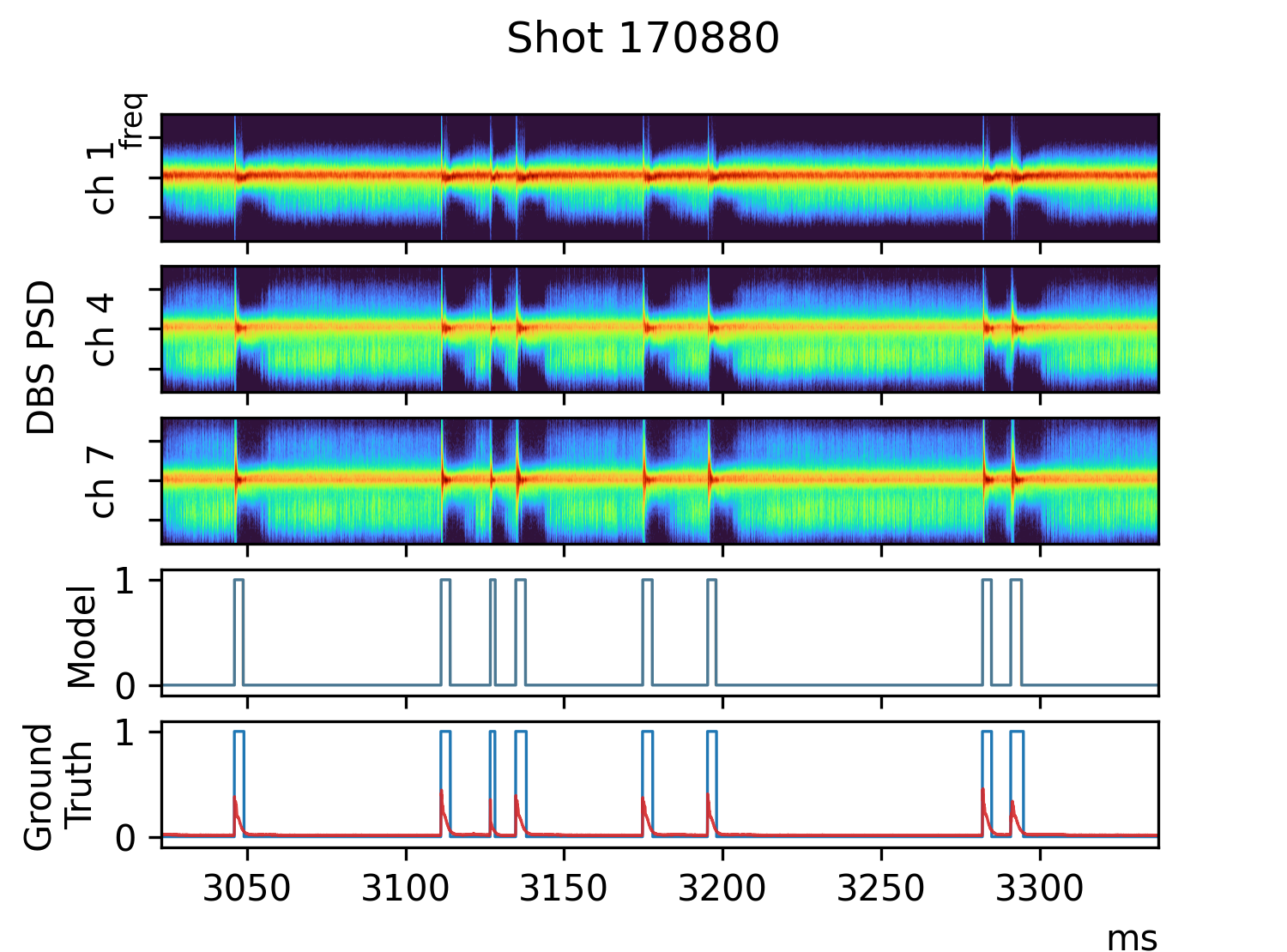}
    \caption{}
    \label{fig:testing_170880}
\end{subfigure}
\hfill
\begin{subfigure}{0.9\columnwidth}
    \includegraphics[width=\textwidth]{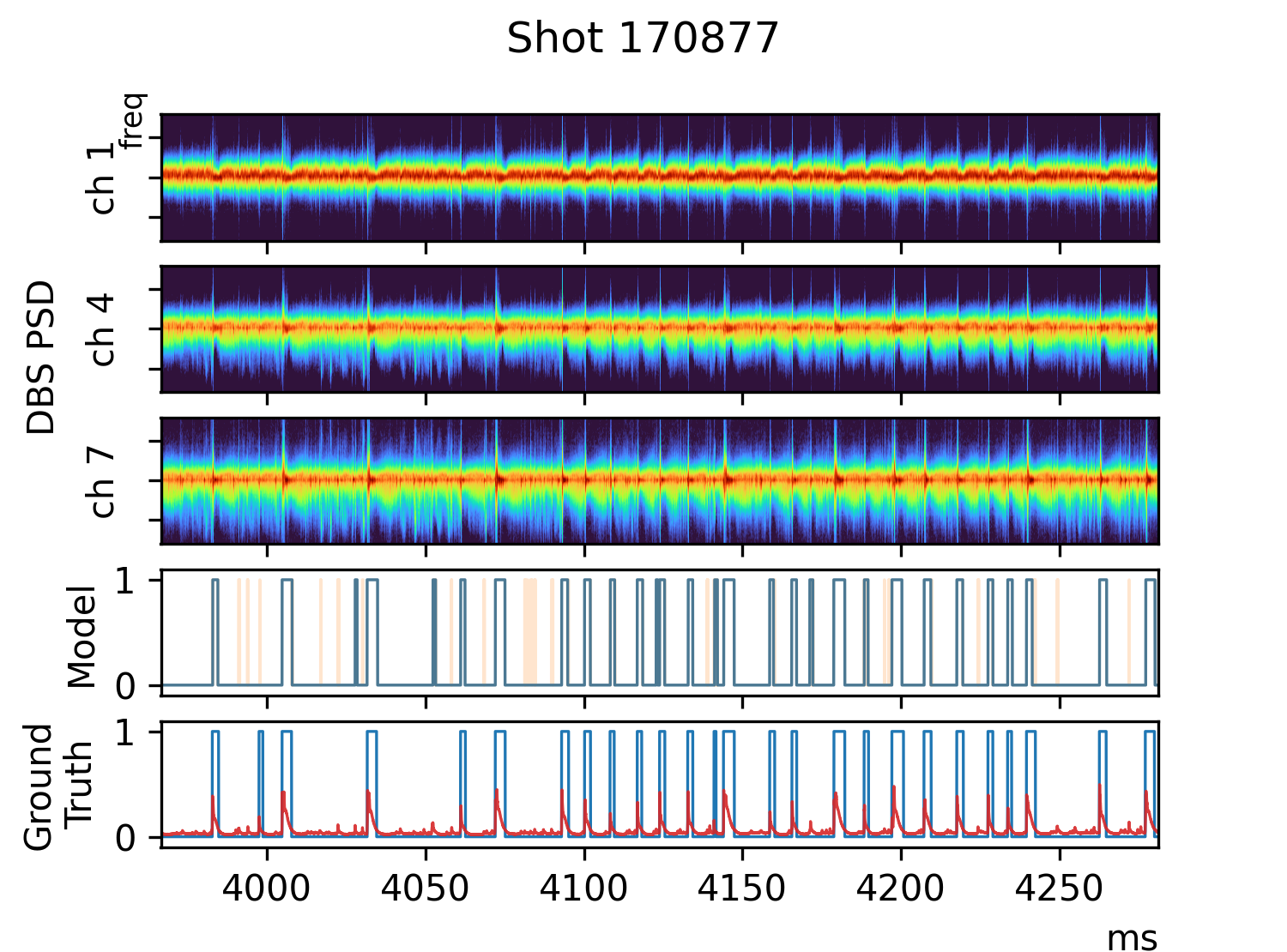}
    \caption{}
    \label{fig:testing_170877}
\end{subfigure}
\caption{Testing on shots similar to the training shots yielded excellent results. The plots above show a representative section covering 10\% of the entire shot. The top three subplots show the DBS PSD data with Doppler shifted frequency as the y-axis and time as the x-axis. The subplot directly below shows the raw model output in orange and the post-processed output (as described in section \ref{Testing}) in blue. The bottom subplot shows the D$_\alpha$ data in red and the ground truth derived from D$_\alpha$ in blue. The average normalized radial positions ($\rho$) of the DBS measurement locations, as calculated with GENRAY \cite{smirnov2009proceedings}, are as follows --- 0.970, 0.957, and 0.941 for channels 1, 4, and 7, respectively, for shot 170877, and 0.992, 0.979, and 0.967 for channels 1, 4, and 7, respectively, for shot 170880. For shot 170880 in (a), every ELM event is detected, where only the duration of each event differing from the ground truth. The raw and post-processed output are almost identical hence the raw model output is not noticeable. For more frequent ELMs in shot 170877 seen in (b), false negatives and positives do appear occasionally.}
\label{fig:testings}
\end{figure*}

The final model was trained for 23 epochs (training cycles). Two metrics are used to judge the training result: the binary cross entropy loss and the f1-score \cite{alzubaidi2021review}. The f1-score is defined as the harmonic mean of the recall and precision of the model results, ranging from 0, no detection, to 1, perfect detection. The f1 score is an insightful measure of model performance in imbalanced datasets, such as those used in this project (since ELMs are sparse). For context, a random predictor achieves an f1-score around 0.1 to 0.2 on the shots used in this paper. The final test loss and f1-score is 0.03 and 0.955, respectively, and the final validation loss and f1-score is 0.08 and 0.897, respectively. On the training shots, it is unsurprising that the model is capable of reproducing the ground truth well. The f1 score of the processed result is high, above 0.93 for all shots.

The model is capable of reproducing excellent results when tested on shots 170877 and 170880, both achieving f1-scores of 0.93. While the f1-score is an important measure of the model performance, visualization of the model output reveals deeper aspects of the model behaviour. Specific sections of shots are presented in the figures to demonstrate the model behavior as shown in Figure \ref{fig:testings}. Here, a counting algorithm is introduced to count the number of successfully detected ELM events, along with the false positives and negatives. The algorithm checks every ground truth event for an overlapping model detection. Note that this is only insightful when combined with other performance metrics since a model that predicts all 1s will have detected all events according to the algorithm. For shot 170880, representing lower frequency ELMs, all 96 ELM events were detected by the model, only differing with the ground truth by the duration of the ELM event. For shot 170877, where ELM frequency is higher, false positives and negatives occur, albeit infrequently. Of the 153 ELM events, 152 were detected with 11 false positives (7\%) and 1 false negative (0.6\%). The false positives are generally accompanied by a D$_\alpha$ spike that was not sufficiently large to be detected as an ELM event by the OMFIT algorithm. False negatives generally only appear for the processed output, where the raw detection from DBS data is too weak and removed by the post processing algorithm. 

\begin{figure*}
\centering
\begin{subfigure}{0.9\columnwidth}
    \includegraphics[width=\textwidth]{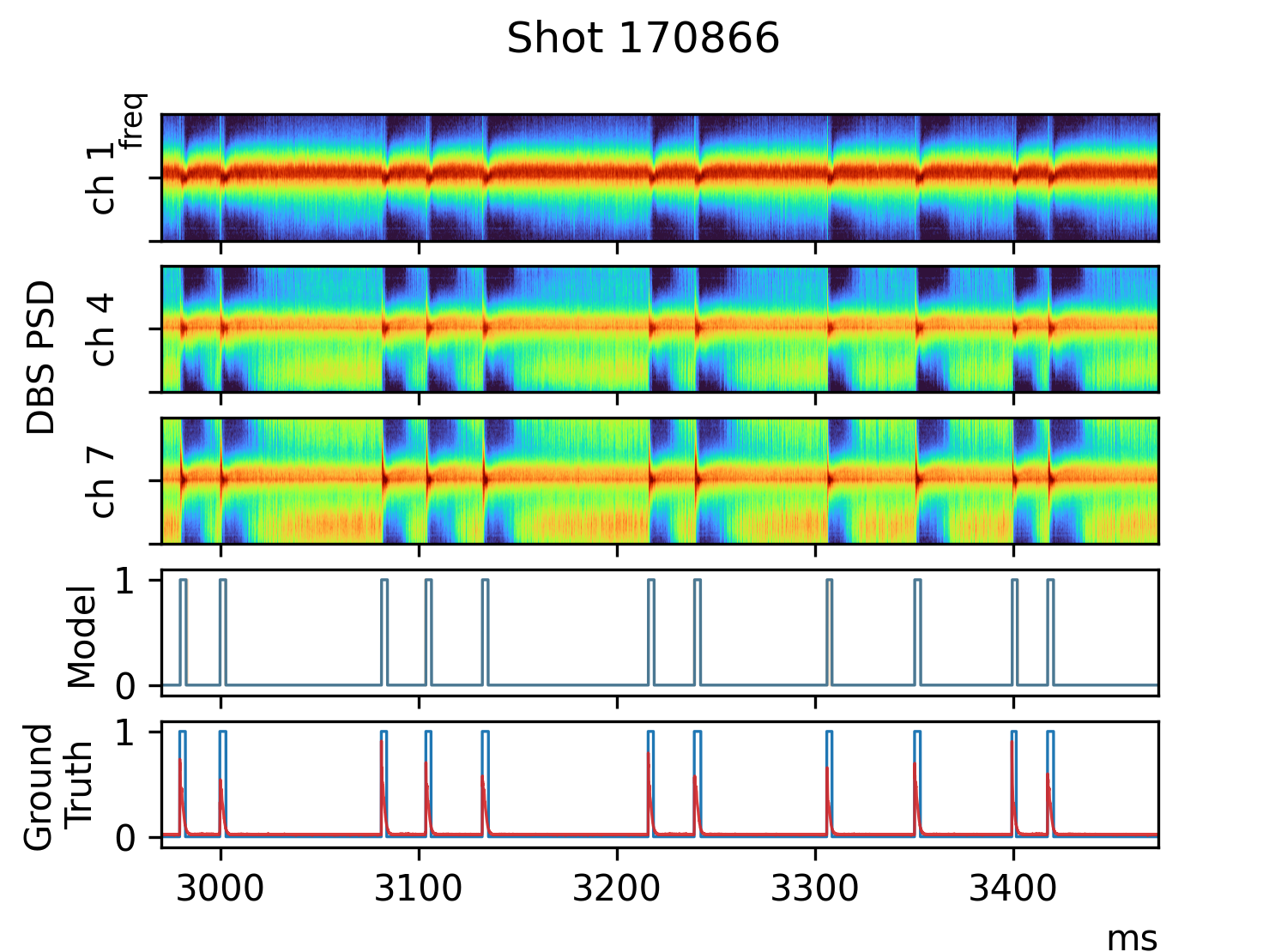}
    \caption{}
    \label{fig:testing_170866}
\end{subfigure}
\begin{subfigure}{0.9\columnwidth}
    \includegraphics[width=\textwidth]{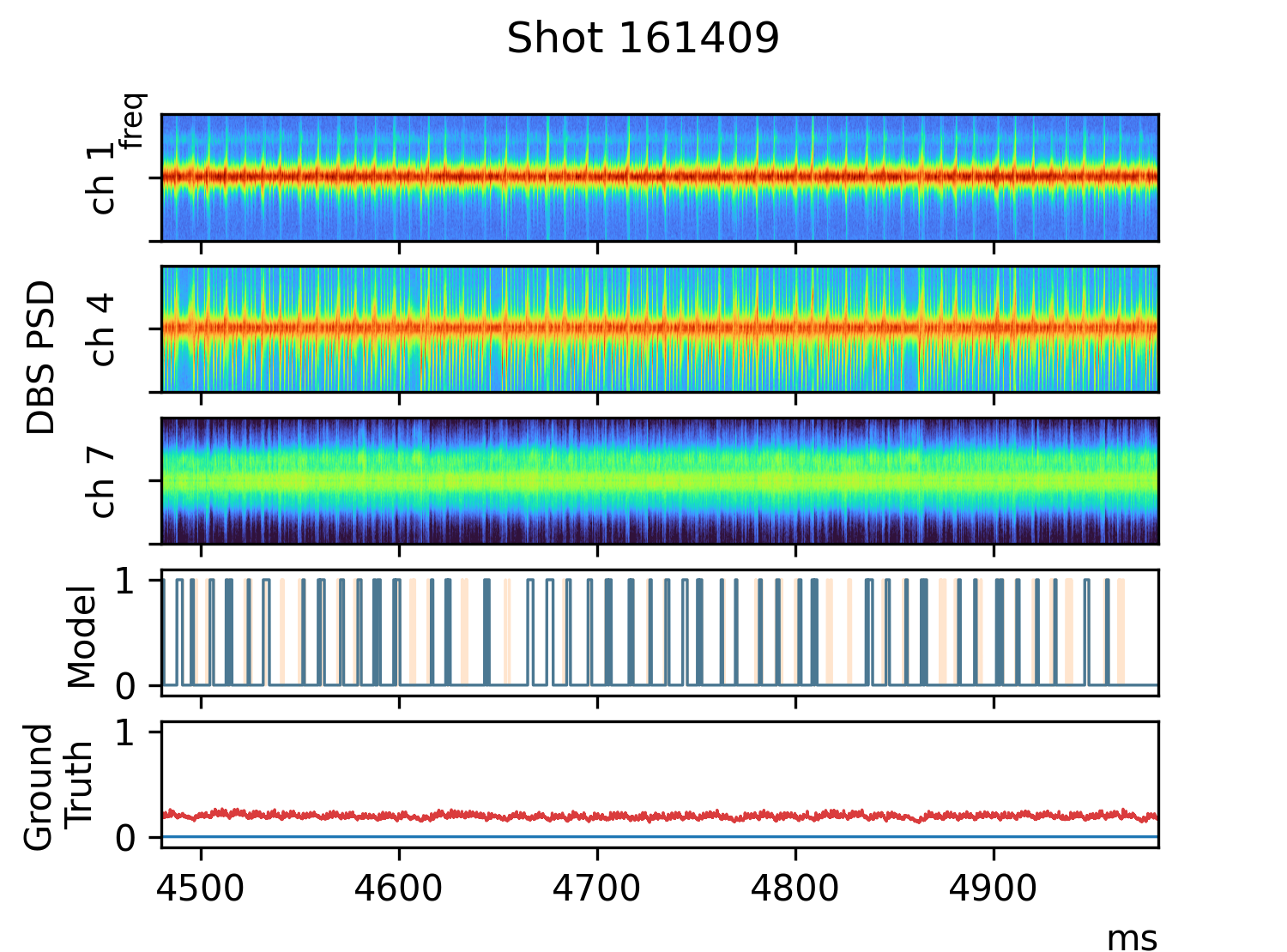}
    \caption{}
    \label{fig:testing_grassy}
\end{subfigure}
\begin{subfigure}{0.9\columnwidth}
    \includegraphics[width=\textwidth]{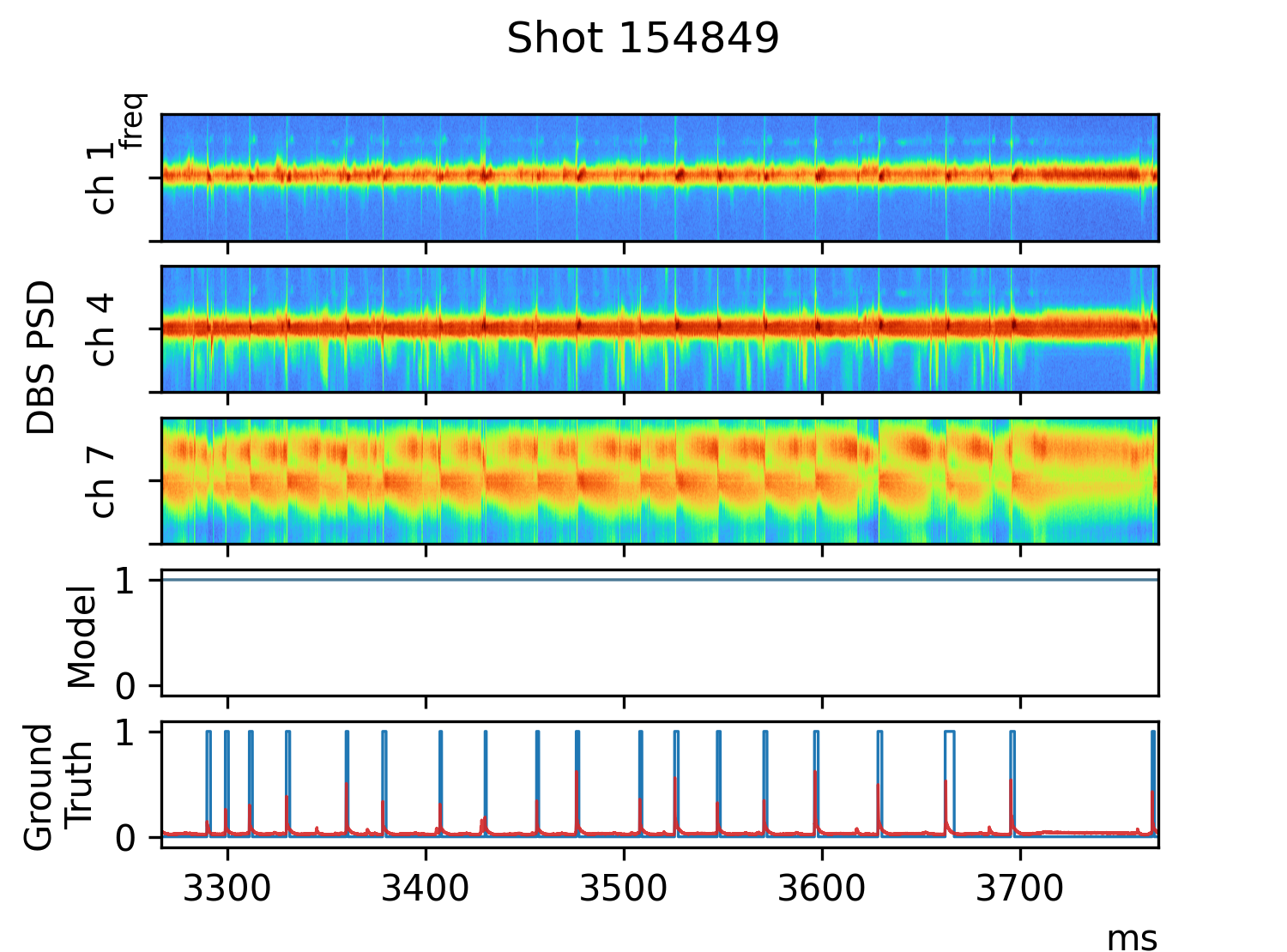}
    \caption{}
    \label{fig:testing_suppressed}
\end{subfigure}
\begin{subfigure}{0.9\columnwidth}
    \includegraphics[width=\textwidth]{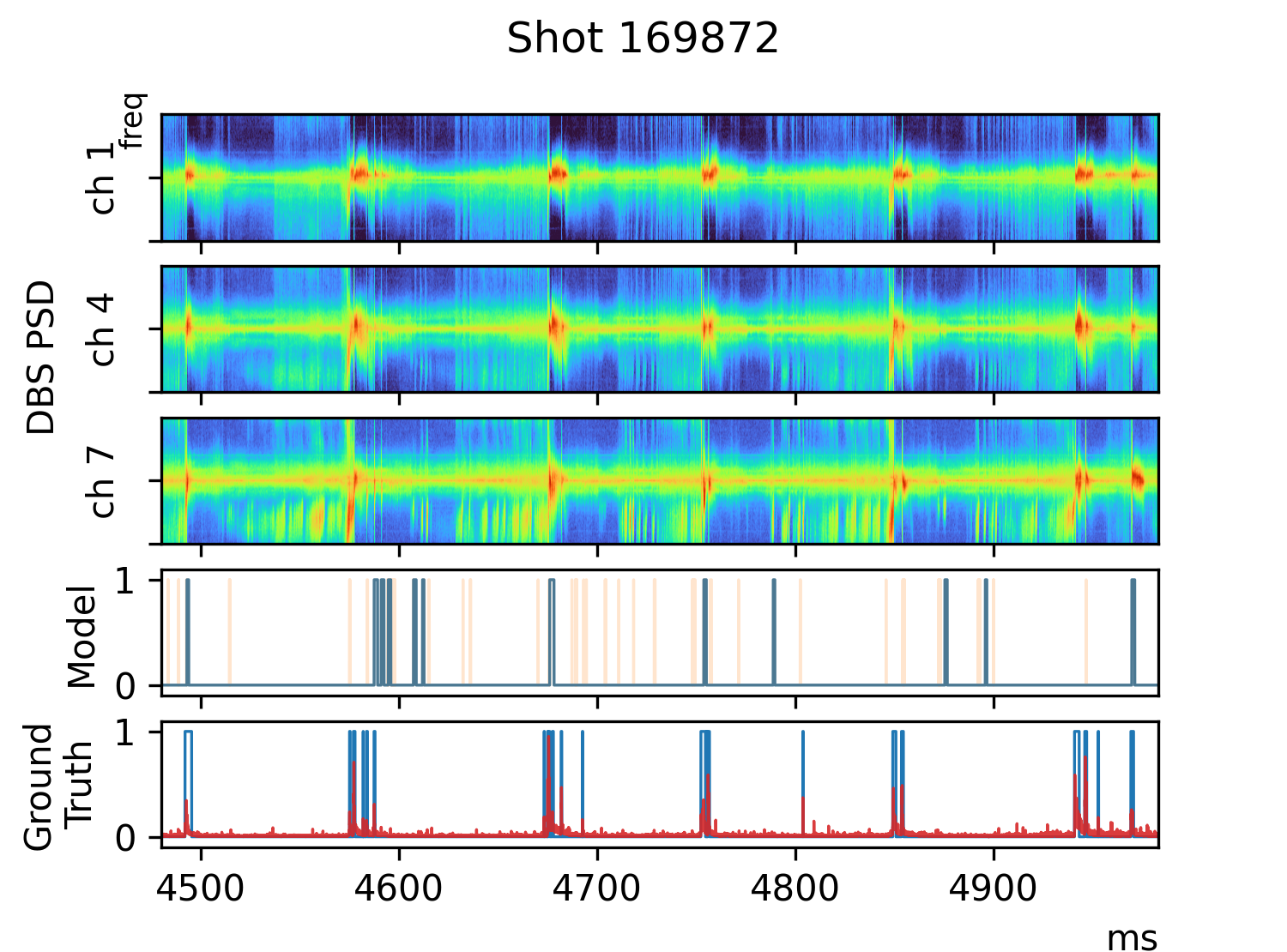}
    \caption{}
    \label{fig:testing_wide}
\end{subfigure}
\caption{Applying our model on shots with different parameters and ELM types. We show a representative section covering 10\% of the entire shot. The top three subplots show the DBS PSD data with Doppler shifted frequency as the y-axis and time as the x-axis. The subplot directly below shows the raw model output in orange and the post-processed output (as described in section \ref{Testing}) in blue. The bottom subplot shows the D$_\alpha$ data in red and the ground truth derived from D$_\alpha$ in blue. The average normalized radial positions ($\rho$) of the DBS measurement locations are as follows --- 0.985, 0.973, and 0.960 for channels 1, 4, and 7, respectively, for shot 170866, 0.945, 0.945, and 0.709 for shot 161409, 0.989, 0.948, and 0.656 for shot 154849, and 0.975, 0.949, and 0.915 for shot 169872 \cite{smirnov2009proceedings}. The model performance is unaffected when the DBS sampling rate used is changed to 5 MHz (a). However, when testing on grassy (b), mitigated (c), and wide pedestal ELMs (d), the model performance is poorer. It should be noted that the D$_\alpha$ data for the shot of grassy ELMs is not a good representative of the ground truth. Here, other diagnostics should be explored to substitute D$_\alpha$ as the ground truth. For mitigated ELMs, the signal from channel 7 deviated from the expected behaviour as the beam cutoff radius penetrates beyond the pedestal during suppression. Hence, it is recommended that either channel 7 not be used as input or the model be trained on mitigated ELMs.}
\label{fig:testings_plus}
\end{figure*}

On the other hand, testing on DBS data with different characteristics had varying degrees of success (Figure \ref{fig:testings_plus}). On shot 180866, which contains type I ELMs recorded at a lower sampling rate of 5 MHz, the model produced similarly excellent performance when compared to the results from shots 170877 and 170880. In the time window of shot 180866 shown in Figure \ref{fig:testing_170866}, the model achieved an f1-score of 0.94. It appears that, within the confines of this study, models trained on a specific sampling rate can be readily applied to data measured at a different sampling rate. The model was also tested on three other types of discharges --- with grassy ELMs (shot 161409), with mitigated ELMs (shot 154849), and with wide-pedestal ELMs (shot 169872). On these shots, model performance is poorer. While the model was capable of reacting to periods of ELM events, it was not capable of replicating the same performance. In the section of shot 169872, shown in Figure \ref{fig:testing_wide}, the model achieved an f1-score of 0.21. It should be noted that the f1-score only applies to the section of the shot shown in Figure \ref{fig:testings_plus} as the shot includes other perturbations that are outside the scope of this study, unfairly decreasing the f1-score of the entire shot. For grassy ELMs in shot 161409, the D$_\alpha$ measurement does not detect any ELM events, making it challenging to ascertain the performance of the model. For shot 154849, shown in Figure \ref{fig:testing_suppressed}, channel 7 produces a signal that is not consistent with channels 1 and 4. Since the higher frequency microwave radiation used in channel 7 probes deeper than the other two channels, this observation can be attributed to perturbations in the deeper layers of the plasma. The normalized DBS beam radius for channel 7 during the shot was 0.656 and was found to be beyond the pedestal. Model performance is greatly affected, detecting an ELM event continuously during suppression. As such, it may be best not to use higher frequency channels in future training and testing. Alternatively, implementing voting ensembles or training on more data may produce satisfactory results.  

\section{Discussion}
There are a few immediate steps to improve the current model. Hyperparameter tuning such as random or grid search will be done to improve model performance. Additionally, we will further train and test the model on other types of ELMs to expand its applicability. For this to be fruitful, careful selection of shots is needed to ensure the D$_\alpha$ data, or any other diagnostic used, is a good representation of the ground truth. In the same vein, a more meticulous approach in choosing OMFIT parameters may produce a more reliable ground truth data for training and testing. Infrared thermography diagnostic should be explored as an alternative for the ground truth for shots containing grassy ELMs since it has been shown to detect grassy ELMs more reliably compared to D$_\alpha$ spectroscopy \cite{ashourvan2019formation}. Another possible path to explore is the implementation of physics-informed components to the model or liquid neural networks to improve model performance. 

Further work will explore models that predict and label ELMs ahead of time. This will likely involve adding LSTMs \cite{van2020review} (Long Short Term Memory) or transformers \cite{soydaner2022attention} to the model. Another area for development is increasing the speed of the model to enable real-time detection. There are various techniques to achieve this, such as sparsification \cite{deng2020model} to reduce the model size.       

The ultimate goal is to create tools for future operational tokamaks to detect and, more crucially, predict ELMs through DBS diagnostics. On a more immediate timescale, this research also has the potential to aid current researchers in data processing tasks. Models can be developed to process large volumes of data or to identify long period magnetohydrodynamic modes that are laborious to find through human observation.

\section{Conclusion}
The results of this study are promising. High f1-scores larger than 0.9 are obtained by the CNN when testing on shots similar to the training shots. The model is also capable of replicating these results on DBS data collected and a different sampling rate.

\begin{acknowledgments}
This work was funded by the Urban and Green Tech Office, A*STAR, Green Seed Fund C231718014, the National Research Foundation, Singapore, by the Nanyang Technological University oversea travels grant, 03INS001464C230OST01, and by JST Moonshot R\&D Grant Number JPMJMS2011. This material is also based upon work supported by the U.S. Department of Energy, Office of Science, Office of Fusion Energy Sciences, using the DIII-D National Fusion Facility, a DOE Office of Science user facility, under Awards DEFC02-04ER54698, DE-SC0019005, and DE-SC0019007. We thank A.A. Schekochihin, N.A. Crocker, V.P. Bui, and Z. Yang for useful discussions. N.Q.X. Teo thanks the Nanyang Technological University (NTU) Singapore, CN Yang Scholars Programme, for financial support.\\ 

\textbf{Disclaimer}. This report was prepared as an account of work partly sponsored by an agency of the United States Government. Neither the United States Government nor any agency thereof, nor any of their employees, makes any warranty, express or implied, or assumes any legal liability or responsibility for the accuracy, completeness, or usefulness of any information, apparatus, product, or process disclosed, or represents that its use would not infringe privately owned rights. Reference herein to any specific commercial product, process, or service by trade name, trademark, manufacturer, or otherwise does not necessarily constitute or imply its endorsement, recommendation, or favoring by the United States Government or any agency thereof. The views and opinions of authors expressed herein do not necessarily state or reflect those of the United States Government or any agency thereof.

\end{acknowledgments}

\bibliographystyle{unsrt}


\end{document}